\documentclass{aa}

\usepackage{natbib}
\bibpunct{(}{)}{;}{a}{}{,} % to follow the A&A style
\usepackage{graphicx}
\usepackage{txfonts}
\usepackage[colorlinks,linktocpage=true]{hyperref}
\usepackage{graphicx}   % Including figure files
\usepackage{amsmath}    % Advanced maths commands
\usepackage{amssymb}    % Extra maths symbols
\usepackage{mathtools}

\begin{document}

\title{Distribution of phantom dark matter in dwarf spheroidals}
% The list of authors, and the short list which is used in the headers.
% If you need two or more lines of authors, add an extra line using \newauthor
%\author[A. O. Hodson et al]{
\author{Alistair O. Hodson
\inst{1,2} \fnmsep\thanks{\email{hodson@to.infn.it}} 
 \and
Antonaldo Diaferio\inst{1,2} \fnmsep\thanks{\email{diaferio@to.infn.it}}
\and 
Luisa Ostorero\inst{1,2} \fnmsep\thanks{\email{ostorero@to.infn.it}} 
}

%\offprints{hodson@to.infn.it, diaferio@to.infn.it, ostorero@to.infn.it}

% List of institutions
\institute {Dipartimento di Fisica, Universit\`a di Torino, Via P. Giuria 1, I-10125 Torino, Italy
\and
Istituto Nazionale di Fisica Nucleare (INFN), Sezione di Torino, Via P. Giuria 1, I-10125 Torino, Italy
}

\date{Received ...; accepted ...}

% Abstract of the paper
\abstract 
{We derive the distribution of the phantom dark matter in the eight classical dwarf galaxies surrounding the Milky Way, under the assumption that modified Newtonian dynamics (MOND) is the correct theory of gravity. According to their observed shape, we model the dwarfs  
as axisymmetric systems, rather than spherical systems, as usually assumed. In addition, as required by the assumption of the MOND framework, we realistically include the external gravitational field of the Milky Way and of the large-scale structure beyond the Local Group. 
For the dwarfs where the external field dominates over the internal gravitational field, the phantom dark matter has, from the star distribution, an offset of $\sim 0.1-0.2$~kpc, depending on the mass-to-light ratio adopted. This offset is a substantial fraction of the dwarf half-mass radius. For Sculptor and Fornax, where the internal and external gravitational fields are comparable, the phantom dark matter distribution appears disturbed with spikes at the locations where the two fields cancel each other; these features have little connection with the distribution of the stars within the dwarfs. Finally, we find that the external field due to the large-scale structure beyond the Local Group has a very minor effect. The features of the phantom dark matter we find represent a genuine prediction of MOND, and could thus falsify this theory of gravity in the version we adopt here if they are not observationally confirmed.
}


\titlerunning{Dwarf Spheroidal Galaxies in QUMOND}  % or: "Dwarf Spheroidals in QUMOND" [no "galaxies" in the full Title]
\authorrunning{A.O. Hodson et al.} 
\maketitle

%%%%%%%%%%%%%%%%%%%%%%%%%%%%%%%%%%%%%%%%%%%%%%%%%%%

%%%%%%%%%%%%%%%%% BODY OF PAPER %%%%%%%%%%%%%%%%%%

\section{Introduction}

If we assume that Newtonian dynamics is applicable on scales of galaxies and beyond, there must be an additional gravitational source, other than baryons, to explain the observed dynamics. This source is most commonly attributed to cold dark matter (CDM) \citep[e.g.][]{DelPopolo2014}. Combining a CDM component with a cosmological constant $\Lambda$ is the essence of the $\Lambda$CDM model, which has been widely adopted \citep[e.g.][]{Ostriker1995,Peebles2015}. 

The $\Lambda$CDM model is most successful in explaining the observed properties of our Universe on cosmic scales (see for example \citealt{Planck2018} and references therein). 
However, in addition to the lack of a direct detection of a dark matter particle and because  the nature of the cosmological constant is still unknown, there are some issues on the galactic scale, for example the missing satellite problem, the cusp-core problem, or the satellite alignments, that make the acceptance of the $\Lambda$CDM model without reservation somewhat difficult \citep[e.g.][]{Famaey2013,Bullock2017,DeMartino2020}. Until these issues can be understood in the context of $\Lambda$CDM \citep[e.g.][]{DelPopolo2017}, investigating alternative scenarios is certainly legitimate. 

One possibility is that  dark matter has different properties to CDM: it could be slightly warmer, self-interacting \citep[e.g.][]{Spergel2000},  very light and fuzzy \citep[e.g.][]{Hu2000,Broadhurst2018}, or it could be a superfluid \citep[][]{Khoury2015}. If a different type of dark matter is not a solution, the observed dynamics on galactic scale might suggest that our gravitational model is incomplete. Most modified gravity theories focus on the nature of the cosmic accelerated expansion rather than on the dynamics of galaxies, and thus still assume the presence of a non-baryonic dark matter component (see e.g. \citealt{Clifton2012} or \citealt{Joyce2015} for reviews). There are fewer modified gravity theories that attempt to remove the necessity of dark matter on galactic
scales. We focus here on Modified Newtonian Dynamics (MOND) \citep{Milgrom1983,Bekenstein1984}, 
which  is the most investigated of these theories \citep{Famaey2012}. 

In this work we investigate the dynamics of dwarf spheroidals in MOND. Revealing  
the nature of dwarf spheroidals is particularly relevant because, in the standard framework, their dynamical properties depend both on the cosmic properties of dark matter and on the nature of the dark matter particle itself \citep[e.g.][]{Broadhurst2019}; more importantly, the current undetection of gamma-ray signals from dark matter particle annihilation \citep[e.g.][]{Strigari2018} intriguingly  vitalises the exploration of theories of gravity with no dark matter. 

The luminosity of the  classical dwarf spheroidals is in the range $\sim 10^{5} - 10^{7}$ L$_{\odot}$ \citep[e.g.][]{Irwin1995,Mateo1998} with velocity dispersion approximately in the range $\sim 6 - 10$ km~s$^{-1}$ \citep[e.g.][]{Walker2007}. In the context of $\Lambda$CDM, by assuming that the dwarf spheroidals are in equilibrium, the Jeans analysis shows that these systems require extended dark matter  halos to explain the velocity dispersion profiles \cite[e.g.][]{Walker2009}. 
\citet{Lokas2009} also suggested that some dwarf spheroidals may be explained by a mass-follows-light model, but they still are dark matter dominated. The main difference between the studies of \cite{Walker2009} and \cite{Lokas2009} is the different method used to identify which stars are members of the dwarf galaxy. As mentioned in \cite{Lokas2009}, her method is much stricter than that of  \citet{Walker2009}. 
Independently of the exact dataset   assumed to construct the velocity dispersion profile, both studies conclude that dark matter must dominate the baryonic matter in these galaxies, although, if     the assumption of dynamical equilibrium is dropped, the estimated amount of  dark matter in dwarf spheroidals might  actually be smaller than currently thought or even totally absent \citep[][]{Hammer2018}.  

Modified Newtonian dynamics assumes that Newtonian gravity breaks down in environments where the gravitational acceleration is less than $\approx 10^{-10}$ m s$^{-2}$. Thus, the dynamics of the dwarf spheroidals are qualitatively understood as they have low internal accelerations, and the deviation from Newtonian gravity should be large. High mass-to-light ratios were indeed predicted by \citet{Milgrom1983} years before they were actually measured \citep[e.g.][]{Mateo1991}. By assuming dynamical equilibrium and spherical symmetry and by adopting a membership identification based on kinematic information, \cite{Serra2010} improved on the work of \citet{Angus2008} to show that MOND is successful in explaining the velocity dispersion profiles of dwarf spheroidals, except for Carina, where the stellar mass-to-light ratio required to match the velocity dispersion data is quite high compared to what is expected for the stellar population. Detailed N-body simulations in MOND do not seem to alleviate this tension \citep{Angus2014}. 

For simplicity, spherical symmetry is often assumed for the stellar distribution of dwarf spheroidals, even though these systems tend to be slightly flattened, with observed minor-to-major axis ratios $\lesssim 0.7$ \citep{Irwin1995}. 
In Newtonian gravity, dwarf spheroidals have been studied by \citet{Hayashi2015} under the assumption that both the stellar and dark matter density components are axisymmetric. This study performed a best-fit Jeans analysis by fitting derived velocity dispersion profiles along three axes, the major, minor, and   intermediate axes, where the intermediate axis   forms an angle of $45^{\circ}$ from the observed major axis. \citet{Hayashi2015} find that the profiles are best fit by a dark matter distribution that is also flattened. However, the flattening predicted for the dark matter is not necessarily the same as that of the stellar component, similarly to  high-resolution simulations of dwarf galaxies from the FIRE project \citep{Gonzalez2017},  which show that the long-to-short axis ratios for the dark matter and stellar components are not necessarily equal.

In this work we improve over previous models of the dwarfs in MOND by self-consistently taking into account two crucial 
ingredients: (1) the asphericity of the stellar distribution, and (2) the external gravitational field acting on the dwarf. 
This external field effect (EFE) is peculiar to MOND: unlike Newtonian gravity, where tidal effects disappear
in systems embedded in a constant external gravitational field, in MOND any external gravitational field across the system
affects its internal dynamics. 

The EFE plays an important role when it comes to explaining  declining rotation curves \citep{Haghi2016} or how satellite systems close to the host galaxy may exhibit Newtonian-like behaviour despite having low internal accelerations \citep[e.g.][]{Famaey2018}. Sometimes, it is not possible to know the exact magnitude of the external field effect and therefore some assumptions must be made. For example, in the work of \cite{Haghi2016} the external field strength was left as a free parameter and then checked to see if it could be physically justified. 

As we   show below, interpreting the Poisson equation in MOND with Newtonian gravity returns the so-called phantom dark matter density.
The asphericity of the stellar distribution and the EFE  thus have two important effects on 
the phantom dark matter halo: its asphericity
and its offset from the stellar distribution \citep{Knebe2009, Wu2010}.

In this work, we investigate these issues in detail. We determine the phantom dark matter distributions of the eight classical dwarf spheroidal systems by adopting the QUasi-linear MOND (QUMOND) formulation \citep{Milgrom2010}. We perform calculations assuming both a spherical and an axisymmetric baryonic matter distribution. We also quantify by how much, if at all, the phantom dark matter peaks are displaced from the baryons.

In Sect.~\ref{QUMONDSec} we outline the MOND equations used to calculate the phantom dark matter density. We briefly discuss some theoretical implications of the MOND paradigm in Sect.~\ref{TheorySec}.  In Sect.~\ref{BaryonsSec} we describe the baryonic model used to  describe the dwarf galaxy and the Milky Way. In Sect.~\ref{PhantomSec} we determine the phantom dark matter density profiles and investigate the relevance of external fields, other than the Milky Way, on the farthest dwarf spheroidals. Finally, we discuss our results in Sect.~\ref{DiscussionSec}  and conclude in Sect.~\ref{ConclusionsSec}.

\section{Quasi-linear MOND}\label{QUMONDSec}

Throughout this work, we will be using the QUMOND formulation \citep{Milgrom2010} rather than  the original  A QUadratic Lagrangian (AQUAL) formulation \citep{Bekenstein1984}. The AQUAL and QUMOND fomulations are identical for spherical systems, but can produce different dynamics for systems of less symmetry. Therefore, our analysis of the dwarf spheroidals is technically testing QUMOND, though the general results and conclusions will also apply to AQUAL. 

The advantage of QUMOND is that the phantom dark matter density can be determined analytically, even for triaxial systems, if the Newtonian gravitational potential is known analytically. The models of dwarf spheroidals we present have an  analytic solution for the Newtonian gravitational potential, making the calculation for the phantom dark matter density simple. 

In QUMOND, the total gravitational potential  $\Phi$ is related to the Newtonian potential $\Phi_{\rm N}$ by the equation
\begin{equation}\label{QUMONDPoiss}
\nabla^2\Phi = \nabla \cdot \left[  \nu(y) \nabla \Phi_{\rm N} \right] \, ,
\end{equation}
where $\Phi_{\rm N}$ satisfies the standard Newtonian Poisson equation
\begin{equation}\label{PoissonN}
\nabla^2\Phi_{\rm N} = 4\pi G \rho \, ,
\end{equation}
with $\rho$ the mass density that we only associate with stars, and $\nu(y)$ the QUMOND interpolation function
\begin{equation}\label{interp}
\nu(y) = \frac{1}{2} + \frac{1}{2}\sqrt{1 + \frac{4}{y} }\, ,
\end{equation}
with $y\equiv |\nabla\Phi_{\rm N}|/a_0$, and $a_0$ the MOND acceleration constant. We adopt the common value $a_0 = 1.2 \times 10^{-10}$ ms$^{-2}$, although slightly different values of $a_0$ associated with  different interpolation functions appear in the literature \cite[see e.g.][]{Hees2016}. Equation\ (\ref{interp}) is usually referred to as the `simple' form of the interpolation function \citep[][]{Famaey2005,Zhao2006}. 

If we interpret Eq.\ (\ref{QUMONDPoiss}) as a standard Poisson equation, the MOND potential
is generated by a source function $\hat\rho = \nabla^2\Phi/4\pi G $. 
With a Newtonian approach, the quantity
\begin{equation}
\rho_{\rm ph} = \hat\rho  - \rho 
\end{equation}
is thus interpreted as the density of the dark matter. In MOND, $\rho_{\rm ph}$ originates from the incorrect interpretation
of the law of gravity and is referred to as the phantom dark matter density.

To explicitly model the MOND EFE, we decompose the MOND potential into two parts: the internal potential $\Phi_{\rm int}$,  
generated by the stars within the dwarf,  and the external potential $\Phi_{\rm ext}$, generated by the mass surrounding the dwarf. We thus write Eq.\ (\ref{QUMONDPoiss}) as
\begin{equation}\label{PhantomDMEqn}
\nabla^2(\Phi_{\rm int} + \Phi_{\rm ext}) = \nabla \cdot \left[  \nu\left(\frac{|\nabla\Phi_{\rm N} + \nabla\Phi_{\rm ext~ N}|}{a_{0}}\right)( \nabla \Phi_{\rm N}+ \nabla\Phi_{\rm ext~N}) \right]\, .
\end{equation}

Interpreting the dynamics of the dwarf governed by the MOND law with a Newtonian approach would thus imply a phantom dark matter density 
\begin{equation}
\rho_{\rm ph} = \frac{1}{4\pi G}\left[ \nabla^2(\Phi_{\rm int} + \Phi_{\rm ext})  - \nabla^2(\Phi_{\rm N} + \Phi_{\rm ext~N})\right] \, .
\label{eq:rho_ph}
\end{equation}

The Newtonian acceleration $\nabla\Phi_{\rm ext~N}$ due to the mass distribution surrounding the dwarf mostly originates from the baryonic matter density of the Milky Way. This acceleration is approximately constant
within each dwarf galaxy, because the dwarf sizes of a few kiloparsec are much smaller than their distance to the Milky Way centre, in the range $\sim 70-250$ kpc. Therefore, 
the $\nabla^2\Phi_{\rm ext~N}$ term in Eq.\ (\ref{eq:rho_ph}) is close to zero and is in general less than $\nabla^2\Phi_{\rm N}$, which is proportional
to the density of the stars within the dwarf. In addition, in Sect.~\ref{BaryonsSec}, we will see that the internal and external Newtonian fields, $\nabla\Phi_{\rm N}$ and $\nabla\Phi_{\rm ext~ N}$, are much smaller than $a_0$. Therefore, the interpolation function $\nu$ in Eq.\ (\ref{interp}) is boosted and, in turn, according
to Eq.\ (\ref{PhantomDMEqn}), $\nabla^2(\Phi_{\rm int} + \Phi_{\rm ext})$ is much larger than $\nabla^2\Phi_{\rm N}$;  
from this argument we expect a large value for the phantom dark matter.  This expectation is confirmed by the observed line-of-sight velocity dispersion profile: assuming
dynamical equilibrium in Newtonian gravity, this profile does indeed imply a large dark matter component \cite[e.g.][]{Walker2009}, as originally predicted by \citet{Milgrom1983b}.

\section{Peculiar properties of the phantom dark matter}\label{TheorySec}

Before moving onto modelling the dwarf spheroidals, we   highlight here some counter-intuitive properties of the phantom dark matter, mostly due to the EFE, an effect lacking in standard Newtonian gravity. Specifically, (1) the mass associated with the phantom dark matter can be negative; (2) the phantom dark matter can take a different shape than the baryon distribution; and (3) the phantom dark matter can be offset from the baryon distribution. 
At the end of this section we also mention that the magnitude of the Newtonian gravitational field going to zero can be an issue for MOND. We briefly discuss these issues in turn.

\citet{Milgrom1986} shows how a negative density of phantom dark matter can be required if MOND is the correct theory of gravity and Newtonian gravity is used to interpret the kinematics of realistic non-isolated systems. A similar result should be expected in the system 
of the Milky Way and the Large Magellanic Cloud \citep{Wu2008}. Negative phantom dark matter densities are a neat prediction that is specific to MOND. In principle, a negative mass density could be observationally confirmed with gravitational lensing \citep{Wu2008}. If actually observed, this result would be a major setback for standard dark matter with Newtonian gravity, which does not clearly predict such a phenomenon.

The QUMOND interpolation function depends on the magnitude of the Newtonian gravitational acceleration which, in turn, is derived from the distribution of the baryonic matter. Therefore,  if the baryonic matter is not spherically distributed, the derived phantom dark matter distribution will also be non-spherical. Intuitively, if the baryonic matter density is assumed to be axisymmetric (e.g. a flattened spheroid), the phantom dark matter density will also be flattened. This guess is certainly correct for isolated systems. However, as mentioned above, the presence of an external field effect can distort the shape of the phantom dark matter. As a consequence, a spherical distribution of baryonic matter can produce a triaxial distribution of phantom dark matter, and a non-spherical distribution of baryonic matter can further alter the shape of the phantom dark matter.

The external field effect also is responsible for the offset of the phantom dark matter from the baryons \citep{Knebe2009,Wu2010}. Due to the external field direction, in addition to its magnitude, the MOND EFE can alter where the phantom dark matter is predicted to be. Therefore, its distribution does not necessarily exactly mirror the baryon distribution of the system within the external field.  We   explain this in more detail for the specific case of dwarf spheroidals in Sect.~\ref{PhantomSec}, but the logic is the same as that applied to other galactic systems.

The MOND paradigm is reliant on the interpolation function. In this work, we mostly assume two-body systems: the dwarf spheroidal and the Milky Way. There will thus be regions between the two galaxies where the total magnitude of the Newtonian gravitational field tends to zero. These regions will be close to the dwarf galaxy whose baryonic mass is much smaller than the Milky Way mass. It is clear from Eq.\ (\ref{interp}) that as $y \rightarrow 0$, $\nu(y) \rightarrow \infty$. This issue with the interpolation function is well known. One possible solution to this problem is to change the interpolation function in such a way that as $y \rightarrow 0$, $\nu(y) \rightarrow 1/\sqrt{y + \epsilon}$, where $\epsilon$ is a small number \citep{Sanders1986,Famaey2007,Famaey2012}. 

\section{Baryonic matter distribution}\label{BaryonsSec}

In this section we illustrate how we model the baryonic matter distribution within the dwarf galaxies (Sect.~\ref{sec:DG}), and the external gravitational
field due to the Milky Way (Sect.~\ref{sec:MW}). We discuss our assumptions in the last three subsections.

\subsection{Model of the dwarf galaxies}
\label{sec:DG}

We base our model of the stellar mass distribution within the dwarf spheroidal galaxies on the Plummer sphere \citep{Plummer1911}
\begin{equation}\label{Plummer}
\rho_*(r) = \frac{3 M_*}{4\pi b^3}\left[1 + \frac{r^2}{b^{2}}  \right]^{-5/2}\; ,
\end{equation}
where $M_*$ is the stellar mass of the dwarf and $b$ is a scale radius, with $\sim 1.3b$ the half-mass radius  \citep[e.g.][]{Walker2009}.

We note that in MOND different forms of the baryon distribution can induce slightly different phantom dark matter profiles. Therefore, choosing a different model for the dwarf spheroidal surface brightness could slightly, quantitatively but not qualitatively, change our results. Here, we are primarily concerned with the distribution of the phantom dark matter as a consequence of the MOND EFE, thus we   leave examining the role of the dwarf baryon distribution for future work.

We derive the mass density profile by multiplying the observed surface brightness profile by a constant stellar mass-to-light ratio $M/L$ in the $V$ band. We explore the values $M/L=1$, $3$, and $5$~$M_{\odot}/L_{\odot}$ that span the range of the mass-to-light ratios found by \cite{Angus2008}. According to the results of stellar population models \citep{Maraston2005},  $5$~$M_\odot/L_{\odot}$ is indeed an upper limit for stellar mass-to-light ratios.     

Throughout this work we adopt the stellar mass and scale radius for the eight classical dwarf spheroidals as given in \citet[][]{McConnachie2012}, who assume a mass-to-light ratio $M/L=1$~$M_\odot/L_{\odot}$ (see Table \ref{DwarfParams}). We scale this stellar mass according to our desired stellar mass-to-light ratio.

\begin{table*}
\caption{Data adopted for the dwarf models. Columns: 1) dwarf name; 2) $V$-band luminosity used to determine the stellar mass; 3) scale length used in the Plummer model for the stellar distribution ($L_V$ and $b$ are from \citealt{McConnachie2012} and references therein); 4--7) coordinates of the Milky Way centre in the Cartesian system where the dwarf is at the origin and the heliocentric distance  (from \citealt{Pawlowski2013,McConnachie2012}; see Sect. \ref{sec:CoordSys}).}
\label{DwarfParams}
\centering    
\begin{tabular}{lllllll}
\hline\hline
Dwarf      & $\rm L_{\rm V}$ ($10^{5}$ L$_{\odot}$) & b (kpc) & $x_{\rm MW}$ (kpc)    & $y_{\rm MW}$ (kpc)   & $z_{\rm MW}$ (kpc) & $r_{\odot}$ (kpc)  \\
\hline
Draco      & 2.7  & 0.196        & -4.3   & 62.2   & 43.2  & $76 \pm 6$\\
Ursa Minor & 2    & 0.28         & -22.2  & 52     & 53.5  &$76\pm3$ \\
Sculptor   & 14   & 0.26         & -5.2   & -9.8   & -85.3  & $86\pm6$\\
Sextans    & 4.1  & 0.682        & -36.7  & -56.9  & 57.8  & $86\pm4$\\
Carina     & 2.4  & 0.241        & -25    & -95.9  & -39.8 & $105 \pm 6$ \\
Fornax     & 140  & 0.668        & -41.3  & -51    & -134.1 & $147\pm 12$\\
        Leo II     & 5.9  & 0.151        & -77.3  & -58.3  & 215.2 & $233\pm 14$\\
Leo I      & 34   & 0.246        & -123.6 & -119.3 & 191.7  & $254\pm15$\\
\hline
\end{tabular}
\end{table*}

As mentioned, dwarf spheroidals are slightly flattened objects. To model this flattening we introduce the parameter $q$, the ratio of the intrinsic semi-minor axis to the intrinsic semi-major axis, to obtain the axisymmetric distribution \citep[see e.g.][]{Hayashi2015}
\begin{equation}\label{rhodwarf}
 \rho_*(m^2) = \frac{3 M_*}{4\pi b^3q}\left[1 + \frac{m^2}{b^2}  \right]^{-5/2},
\end{equation}
where $m^2 \equiv R^2 + z^2/q^2$; $R$ and $z$  are the coordinates in the frame of the dwarf galaxy such that the flattening is in the $z$-direction; $q$ is related to the intrinsic ellipticity, $\epsilon_{\rm int}$, via 
\begin{equation}
        q^2 = 1 - \epsilon_{\rm int}^2 \, ;
\end{equation}
 and $M_*$ is the total stellar mass.

The next step is to calculate the Newtonian gravitational potential $\Phi_{\rm N}$ that appears in Eq. (\ref{eq:rho_ph}) for systems described by the density distribution of Eq.\ (\ref{rhodwarf}). According to Eq.\ (2.125b) in \citet{Binney2008}, with their arbitrary constant $a_0 $ set to $1$ (but unimportant here),  we have the Newtonian gravitational potential due to the stellar distribution within the dwarf
\begin{equation}
 {\Phi_{\rm N}(R,z) \over 2\pi G } = 
 -  {\gamma\cot\gamma\Psi(\infty) - \frac{q}{2} \int^{\infty}_0 \frac{\Psi(\tilde{m})}{(\tau + 1)\sqrt{\tau + q^2}} \, {\rm d}\tau }\, ,
\end{equation}
where $q^2=\cos^2\gamma$, 
\begin{equation}
 \tilde{m^2} = \frac{R^2}{\tau + 1} + \frac{z^2}{\tau + q^2},
\end{equation}
and
\begin{equation}
\Psi(m) = \int^{m^2}_{0} \rho_*(m^{2}) \, {\rm d}m^{2} = {M_*\over 2\pi b} 
\left[ 1 -\left(1+{m^2\over b^2}\right)^{-3/2}\right]\, ,
\end{equation}
for our model of the ellipsoidal Plummer distribution.

\subsection{External field originated by the Milky Way}
\label{sec:MW}

As mentioned previously, when modelling systems in MOND, the EFE must be taken into consideration. We choose to do this in a self-consistent manner by including a simple potential for the Milky Way. As we are looking beyond spherical symmetry, the advantage of handling the external field in this manner 
is that we automatically get an external field with   $x$, $y$, and $z$ components, so we can place a dwarf spheroidal at its proper location with respect to the Milky Way, and thus provide a more realistic prediction of the expected distribution of the phantom dark matter. 

We model the Milky Way as a single Miyamoto-Nagai stellar disk 
yielding a gravitational potential  
\citep{Miyamoto1975}
\begin{equation}\label{MNModel}
\Phi_{\rm MW}(x,y,z) = -\frac{G M_{\rm MW}}{\sqrt{x^2 + y^2 + \left( a_{\rm d} + \sqrt{z^2 + b_{\rm d}^2}\,  \right)^2}}\, , 
\end{equation}
where $z$ is the dimension perpendicular to the plane $(x,y)$ of the disk, $M_{\rm MW}$ is the baryonic mass of the Milky Way, and $a_{\rm d}$ and $b_{\rm d}$ are scale lengths. We list the values for these three parameters in Table \ref{parameters}.

We adopt the Miyamoto-Nagai disk for the Milky Way model because it provides a fully analytic expression for the Newtonian potential,
unlike exponential disk models, that actually describe real disk galaxies more accurately (see e.g. \citealt[][]{McGaugh2008} for a Milky Way mass model in the context of MOND). A more realistic model of the Milky Way would include a gas disk and a bulge in addition to the stellar disk we only consider here. We can safely ignore the presence of a bulge because it introduces negligble corrections to the potential originated by the Milky Way at the large distances of the dwarfs. 

For the effect of the gas disk we consider the following argument. Recently, to calculate the escape velocity from the Milky Way in the context of MOND, \cite{Banik2018} adopt the baryon distribution of the Milky Way
\begin{equation}\label{ExpModel}
\rho_{\rm b} = \rho_{\rm *~disk} + \rho_{\rm gas~disk} + \rho_{\rm gas~corona}\, ,
\end{equation}
where the three density terms refer to contributions from the stellar disk, gas disk, and gaseous corona, respectively. Again, the bulge was ignored as they are interested, as we are, in the field far away from the Milky Way centre, where the bulge is not dominant. 

To show that our model of the Milky Way with a Miyamoto-Nagai disk is appropriate for our problem, we compare the Newtonian external field $a_{\rm MN}$ produced by our Miyamoto-Nagai model, at the position of the dwarf spheroidals, with the Newtonian external field $a_{\rm Exp}$ estimated with a model similar to the model used in \cite{Banik2018}. Specifically, we neglect the density contribution from the gas corona and set $\rho_{\rm gas~corona}=0$, and, for both $\rho_{\rm *~disk}$ and $\rho_{\rm gas~disk}$,  we adopt the form
\begin{equation}\label{ExpProfile}
\rho_{\rm Exp} = \rho_0 {\rm sech} \left(- \frac{z}{2z_0} \right)\exp\left(- \frac{R}{R_0} \right)\; ,
\end{equation}
where the values of $\rho_0$ for the star and gas disks were calculated assuming the nominal disk masses in \cite{Banik2018}, and the scale radii $R_0$ for both disks were taken from \cite{Banik2018}. We adopt a disk height of $z_0=300$~pc for both disks. Table \ref{parameters} lists the values of the parameters of this Milky Way model.

\begin{figure}
\centering
\includegraphics[scale=0.3]{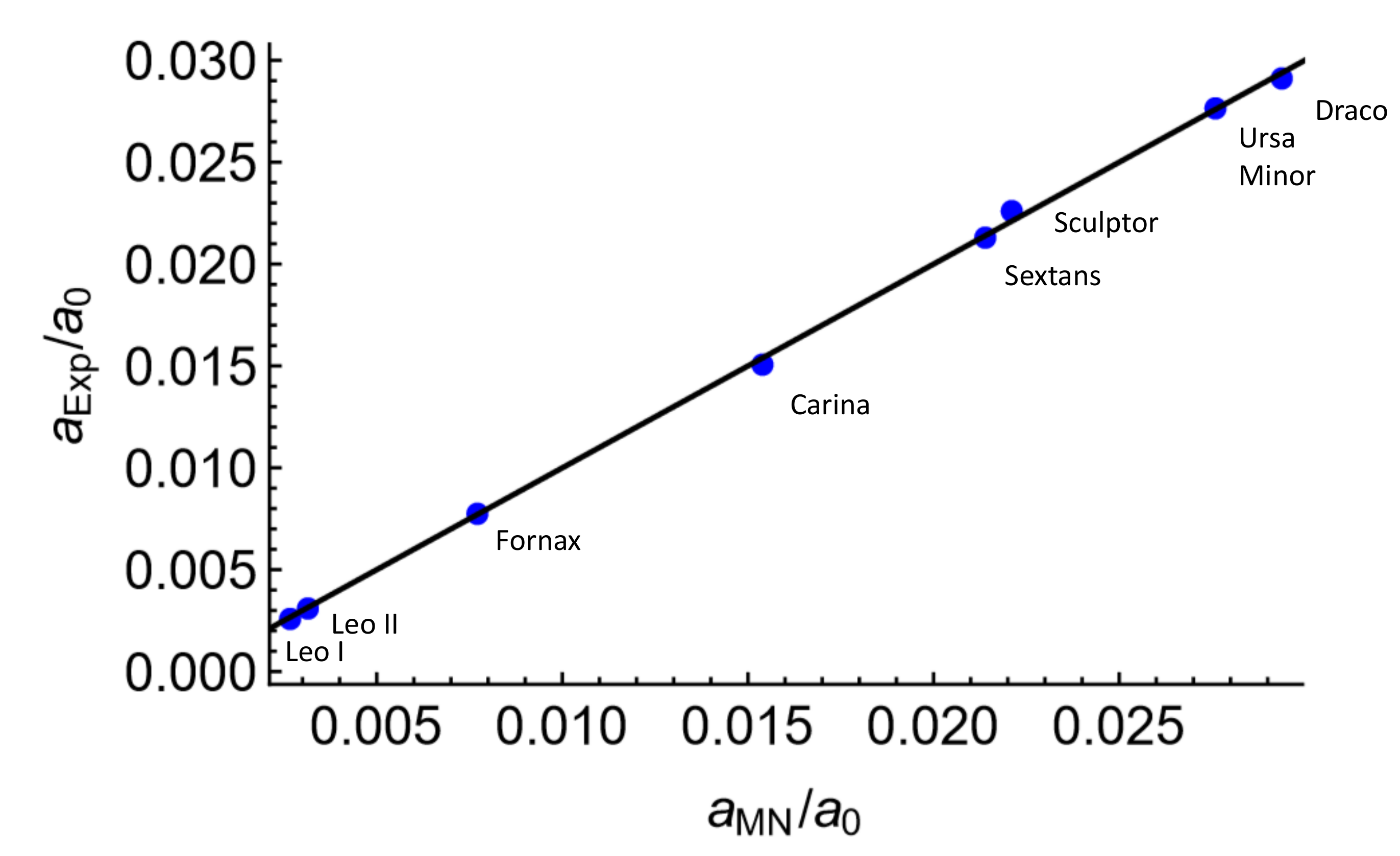}
\caption{Comparison of the magnitude of the Newtonian gravity from two Milky Way models, the Miyamoto-Nagai model $a_{\rm MN}$ (Eq.\ \ref{MNModel}), and a model using exponential disks, $a_{\rm Exp}$, similar to that of \citet{Banik2018} (Eqs.\ \ref{ExpModel} and \ref{ExpProfile}) at the position of the eight dwarf galaxies. The model parameters for the two models are given in Table \ref{parameters}. The black line is a one-to-one line. We see that the two models give similar gravitational strengths.}
\label{ExternalFieldCompare}
\end{figure}

Figure~\ref{ExternalFieldCompare} shows the comparison between the magnitudes of $a_{\rm MN}\equiv|\nabla\Phi_{\rm ext~N}|$ with $\Phi_{\rm ext~N}$ at the position of the dwarf from Eq.\ (\ref{MNModel}), and  $a_{\rm Exp}\equiv|\nabla\Phi_{\rm Exp~N}|$ derived from the solution of the Poisson equation $\nabla^2\Phi_{\rm Exp~N}=4\pi G \rho_{\rm b}$, with $\rho_{\rm b}$ from Eqs.\ (\ref{ExpModel}) and (\ref{ExpProfile}). We see that our Miyamoto-Nagai model gives  values of the magnitude of the Newtonian gravitational field comparable to those of the exponential model. 

Although the magnitude of the gravitational field is similar in the two models, the Cartesian components of $a_{\rm MN}$ and $a_{\rm Exp}$ can be different. Taking Sculptor as an example, the gravitational field due to the Milky Way is mostly oriented along the $z$-axis (see Table \ref{DwarfParams}); therefore, although the $x$ and $y$ components of the gravitational field differ by as much as 7\%, we only see a small difference between the two Milky Way models because the $z$ components in the two Milky Way models  differ by only 3\%. Similarly, for Carina, the largest contribution to the gravitational field strength is in the $y$-direction, and, although the $z$ components differ by approximately 10\%, the $x$ and $y$ components of each Milky Way model are almost identical, with a difference smaller than 1\%; therefore, when we compare the magnitude of the gravitational field strength of the two Milky Way models in Fig.~\ref{ExternalFieldCompare}, we only see a mild difference around 1\%. Similar arguments hold for the remaining dwarfs.  
We therefore conclude that choosing the simpler Miyamoto-Nagai model will not drastically affect our results.

\subsection{Our coordinate system:  dwarf model and  external field combined}
\label{sec:CoordSys}

To properly estimate the EFE due to the Milky Way on each dwarf, we need to locate the two galaxies in a reference frame. 
We assume a Cartesian reference frame where the origin is located at the centre of the dwarf. We define the $z$-axis to be perpendicular to the Milky Way disk with the positive direction pointing towards the galactic north. 
The $x$-axis is parallel to the line that points from the Milky Way centre to the Sun, and the $y$-axis is perpendicular 
to both the $x$- and $z$-axes, and lies on a plane parallel to the Milky Way disk. 
Our choice differs from the \citet{Pawlowski2013} frame where the origin is the Milky Way centre.

In our Cartesian frame, the Newtonian external field acting on the dwarf galaxy is thus 
\begin{align}
        \label{eq:PhiExtN}
\begin{split}
\Phi_{\rm ext~N}(x,y,z) &= \Phi_{\rm MW}(x-x_{\rm MW},y-y_{\rm MW}, z-z_{\rm MW}),
\end{split}
\end{align}
where $x_{\rm MW}$, $y_{\rm MW}$, and $z_{\rm MW}$ are the coordinates of the centre of the Milky Way.  Table \ref{DwarfParams} lists these coordinates adapted from \citet{Pawlowski2013}.

We wish to estimate the profile of the phantom dark matter that, in principle, can be observed from the Sun. We thus adopt the distance of the Sun from the Milky Way centre $R_\odot = 8$~kpc \citep{Boehle2016}, and rotate the dwarf to match its observed shape.
As described in Sect.~\ref{sec:DG}, the equations that we adopt to model the axisymmetric dwarf spheroidals in three dimensions assume that the flattening occurs in the $z$ direction in the dwarf system. Observationally we see the dwarf projected on the sky with an observed ellipticity, whose value is $\epsilon_{\rm obs}\sim 0.3$ for most of the dwarfs. To realistically reproduce the observed situation we thus choose an intrinsic ellipticity $\epsilon_{\rm int}\sim 0.5$, or $\sim 0.7$ (see Sect.~\ref{PhantomSec}  
below), and rotate the dwarf around the major axis until the observable ellipticity matches the actual observed ellipticity. 

\begin{table*}
\caption{Parameters of the Milky Way models. Top half of  table: Parameters  adopted for the Miyamoto-Nagai model (Eq.\ \ref{MNModel}), which we use to  calculate the phantom dark matter profiles for the dwarf spheroidals. Bottom half of  table: Parameters used in \citet{Banik2018} for a realistic MOND Milky Way model, which includes exponential disks (see our Eqs.\ \ref{ExpModel} and \ref{ExpProfile} for details). This second model is  used here only for  comparison to  the Miyamoto-Nagai model (our Fig.~\ref{ExternalFieldCompare}) to show that our model is physically justified. }
\centering
\begin{tabular}{ccc}
\hline\hline
Symbol & Parameter & Value \\
\hline
$M_{\rm MW}$ & disk mass & $1.3 \times 10^{11}$ M$_{\odot}$\\
$a_{\rm d}$ & disk scale length & 5 kpc \\
$b_{\rm d}$ & disk scale length & 0.3 kpc\\
\hline
$M_{*}$ & stellar disk mass & $5.51 \times 10^{10}$  M$_{\odot}$ \\
$M_{\rm gas}$ & gas disk mass& $1.18 \times 10^{10}$ M$_{\odot}$ \\
$R_{\rm 0 *}$ &  stellar disk scale length& 2.15 kpc\\
$R_{\rm 0~gas}$ &  gas disk scale length& 7 kpc\\
$z_{\rm 0*},z_{\rm 0~gas}$ &  stellar and gas disk scale height& 0.3 kpc\\
\hline
\end{tabular}
\label{parameters}
\end{table*}

\subsection{Comparing the external and internal Newtonian fields of the dwarf galaxies}
\label{sec:comparison}

Now that we have set up our model, we can compare the strengths of the Newtonian internal ($a_{\rm int}$) and external ($a_{\rm ext}$) fields. The external field  acting on the dwarf is approximately constant. We compare the value of the external to the internal Newtonian field of the dwarf at the scale radius, $b$, of the Plummer model (Eqs.\ \ref{Plummer} and \ref{rhodwarf}). For simplicity, here we calculate the internal field at this radius assuming spherical symmetry ($q=1$). 

\begin{figure}
\centering
\includegraphics[scale=0.4]{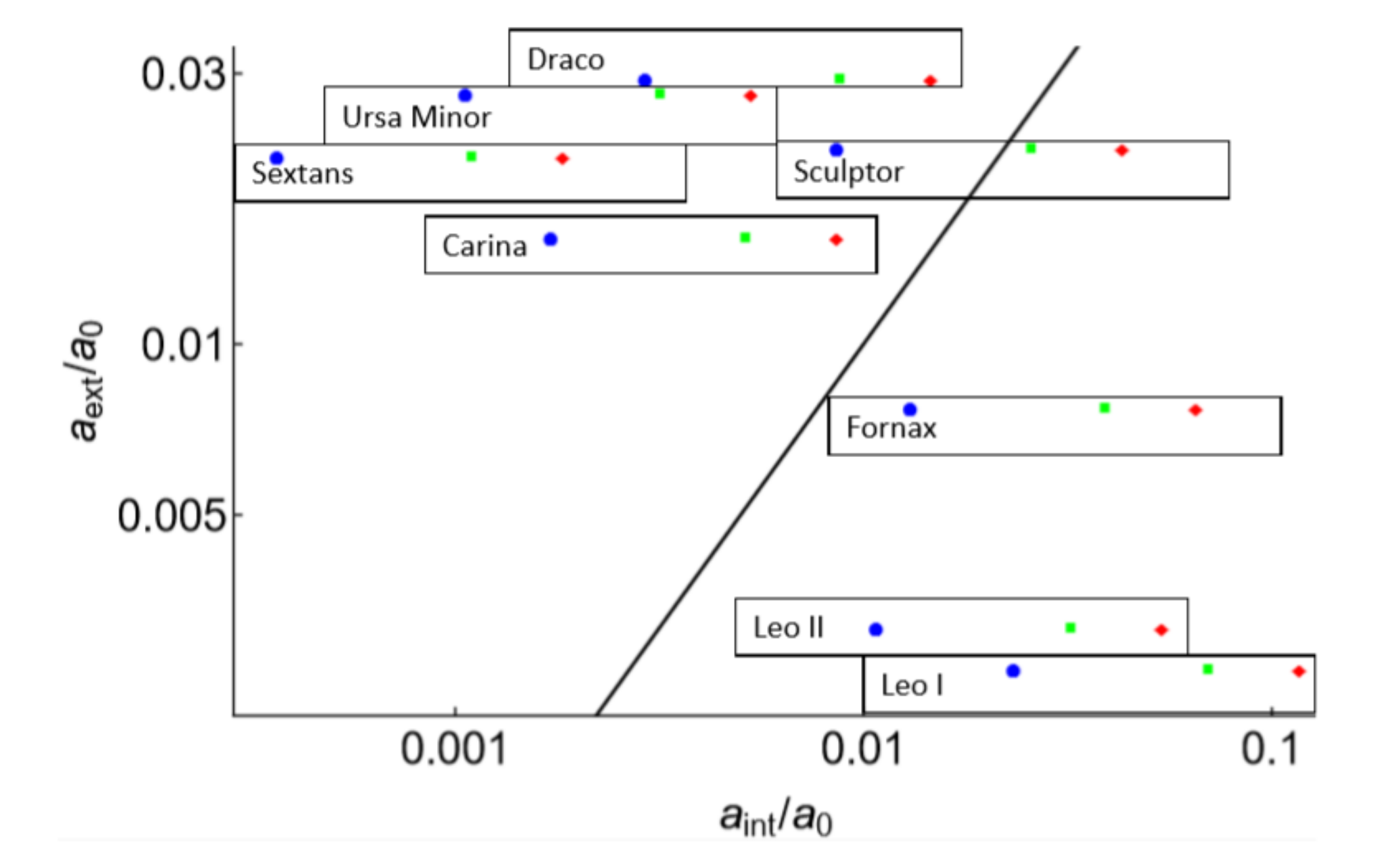}
\caption{Comparison of the internal Newtonian gravity of each dwarf at the scale radius $b$ to the Newtonian external field from the Milky Way at the position of the dwarf.  Blue, green, and red dots represent the internal gravitational strength calculated assuming a stellar mass-to-light ratio of 1, 3, and 5 $M_\odot/L_\odot$, respectively.  The black line is a one-to-one line added to allow easy comparison.}
\label{InternalFieldCompare}
\end{figure}

Figure~\ref{InternalFieldCompare} shows this comparison. The blue, green, and red dots in this figure represent the Newtonian gravitational strength for the dwarf spheroidals given a stellar mass-to-light ratio $M/L=1$, $3$, and $5$~$M_\odot/L_\odot$, respectively.  The black line is a one-to-one line added to allow easy comparison. In this figure we show the Newtonian contribution to gravity, not the total MOND gravitational acceleration.

Each dwarf appears with three points, which are horizontally aligned with the same $a_{\rm ext}$ coordinate. From top to bottom we have Draco, Ursa Minor, Sculptor, Sextans, Carina, Fornax, Leo II, and Leo I. The closer the dwarf is to the Milky Way centre, the stronger the external field. This correlation does not need to be true, in general, given that the Milky Way is not spherical, but it   is  true for these eight systems.  Increasing mass-to-light ratios yield   
increasing $a_{\rm int}$, as expected. 

Figure~\ref{InternalFieldCompare} clearly shows that for the dwarfs Draco, Ursa Minor, Sextans, and Carina, which are located in the top left corner for low stellar mass-to-light ratios, the EFE is expected to be substantially more relevant than for the dwarfs Leo I and Leo II, located in the bottom right corner for high stellar mass-to-light ratios.
For the two remaining dwarf galaxies, Sculptor and Fornax, the internal and external fields are comparable. 

Figure~\ref{InternalFieldCompare} only shows one value of the internal field, at the scale radius $b$, 
which is  not representative of the whole galaxy. Far away from the centre of the dwarf, the internal field can be much weaker than this value. Therefore, Carina, which is just left of the one-to-one line in Fig.~\ref{InternalFieldCompare}, is dominated by the external field in its outer regions, whereas 
Fornax, which is just to the right of the one-to-one line, has an internal field comparable to the external field in its outer regions. In Sect.~\ref{PhantomSec} we   discuss these cases in detail.

\subsection{External field originated by the large-scale structure beyond the Milky Way}\label{LargeScaleSec}

We   assume that the external field acting on the dwarf galaxy is dominated by the Milky Way. However, there are other sources of external field and here we discuss how relevant these sources are. We consider the Andromeda galaxy and the large-scale structure farther away. 

Assuming that Andromeda has a baryonic mass of $M_{\rm A}\sim (1-2) \times 10^{11}$ M$_\odot$ at a distance of $d \sim 800$~kpc from the Milky Way \cite[e.g.][]{Tamm2012}, its Newtonian contribution to the external field is $GM_{\rm A}/(d^2 a_0)\sim 0.0002  - 0.0004 $.  This contribution is one to two orders of magnitude smaller 
than that of the Milky Way $a_{\rm MW}/a_0\sim 0.003-0.03 $ (see Fig.~\ref{ExternalFieldCompare}). 

In terms of an external field from the large-scale structure beyond the Local Group, recent constraints from Milky Way escape velocity measurements require a total external field, acting on the Milky Way itself, around $0.03 a_0$  in the MOND framework \citep{Banik2018}. We can crudely approximate the Newtonian contribution to this external MOND field by solving the curl-free, spherical solution to Eq.\ (\ref{QUMONDPoiss}),
\begin{equation}
a_{\rm ext} = \nu(a_{\rm N~ext}/a_0) a_{\rm N~ext},
        \label{eq:aextLG}
\end{equation}
for $a_{\rm N~ext}$. Doing this yields  $a_{\rm N~ext}/a_0 \approx 0.0009$. If we assume that this field remains roughly constant within a few hundred kiloparsec from the Milky Way, we see that the   dwarf spheroidals that are farthest away from the Milky
Way (Leo I and Leo II, at $\sim 254$ and $\sim 233$ kpc, respectively)  are affected by a Newtonian external field originating from structures beyond the Local Group which is around one-third of the Newtonian external field $a_{\rm MW}/a_0\sim 0.003$ (Fig.~\ref{ExternalFieldCompare}) originating from the Milky Way itself. Therefore, at least for  Leo I and Leo II, the external fields due to the Milky Way and to the structures beyond the Local Group become comparable, and the field due to structures beyond the Local Group should in principle be included in our models.

Clearly, in addition to the estimate of the magnitude of the external field strength from the large-scale structure, we need to determine its direction. From \cite{Wu2008}, who model the Large Magellanic Cloud in MOND, the main source of the external field is assumed to originate from the Great Attractor region, which is in the Sun-Galactic centre direction. 
We adopt the same model. We  discuss exactly how we include this external field in our model in Sect.~\ref{PhantomSec}. 

For the sake of completeness, we note   that \cite{Wu2008} assume a slightly weaker  external field acting on the Milky Way.  They estimated a total external field from beyond the Local Group of $0.01 a_0$, rather than $0.03 a_0$ as estimated by \cite{Banik2018}. Using the curl-free MOND equation, this  $0.01 a_0$ external field  translates to a Newtonian field of $\sim$ 0.0001 $a_0$.  This value is comparible to the value for the Newtonian field acting on the Milky Way from Andromeda ($\sim 0.0002  - 0.0004 a_0$). We choose to keep the stronger external field strength $0.03 a_0$, in line with \cite{Banik2018}, but we caution that if the total external field is indeed weaker, it can become comparable to the field of Andromeda. In this case, the field of Andromeda will need   
to be included in any calculation. 
In Sect.~\ref{PhantomSec}, we  first perform the analysis without the EFE from the large-scale structure beyond the Local Group, then we re-introduce it for Leo II as a case study. 

Hypothetically, if the  total internal gravitational potential of a far dwarf galaxy is known accurately, by making a detailed mass model of the Local Group it could be possible to estimate the external field across the dwarf due to the mass distribution beyond the Local Group, and test whether this contribution is consistent with the estimate of \citet{Banik2018}.

\section{Phantom dark matter profiles}\label{PhantomSec}

We show the phantom dark matter density profiles for each of the eight Milky Way dwarf spheroidals  calculated from Eqs.\ (\ref{PhantomDMEqn}) and (\ref{eq:rho_ph}).  In Sect.~\ref{sec:DwarfAndMW} we consider the EFE due to the Milky Way alone, and in Sect.~\ref{sec:LeoII} we include the additional effect of the large-scale structure beyond the Local Group. 

For each dwarf we  show the results assuming a stellar mass-to-light ratio of $1$, $3,$ and $5$~$M_\odot/L_\odot$, to capture the MOND prediction for the lowest and highest stellar mass estimates. For the flattening, we assume an average intrinsic ellipticity $\epsilon_{\rm int} = 0.5$. This value is chosen in line with the analysis of \citet[][]{Salomon2015}, which examined intrinsic axis ratios of satellite galaxies of Andromeda. The only caveat to this assumption is the case of Ursa Minor, whose observed ellipticity is $\epsilon_{\rm obs}=0.56$, much flatter than the other seven dwarf spheroidals, whose observed ellipticity is $\epsilon_{\rm obs}\sim 0.3$. The sample of galaxies from \cite{Salomon2015} contains systems that have  observed ellipticities similar to that of Ursa Minor. The three values of  intrinsic ellipticity estimated by \citet{Salomon2015} for these systems are in the range 0.65 - 0.77. For Ursa Minor we thus adopt the average intrinsic  value $\epsilon_{\rm int}= 0.7$.

We show the three-dimensional distribution of the phantom dark matter derived from our calculation as follows. 
We show the density profile along three directions: the line of sight as observed from the Sun, the observed major axis, and the observed minor axis. These lines are chosen such that they are perpendicular to each other and intersect at the centre of the dwarf galaxy.

Finally, to highlight the deviation from the spherical approximation, we also plot a normalised phantom dark matter density $\tilde{\rho} = \rho/\rho_{\rm sph}$ where $\rho_{\rm sph}$ would be the prediction if we assumed a spherical distribution for the baryonic matter within the dwarf. Clearly, we always use the axisymmetric model for the Milky Way even when calculating $\rho_{\rm sph}$.

\subsection{Dwarfs in the field of the Milky Way}
\label{sec:DwarfAndMW}

\begin{figure*}
\centering
\includegraphics[scale=0.8]{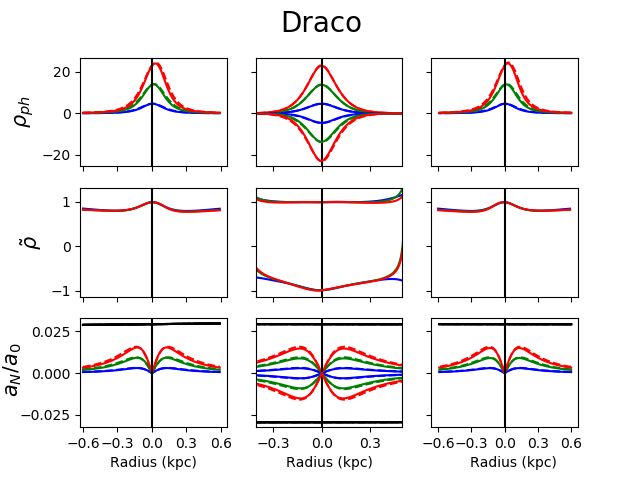}
\caption{Phantom dark matter in Draco. The blue, green, and red curves assume a stellar mass-to-light ratio of 1, 3, and 5 $M_\odot/L_\odot$, respectively. The solid curves are for a flattened baryonic component and the dashed curves are for a spherical distribution.  The panels in the upper row show the distribution of the phantom dark matter in units of $10^7$~M$_\odot$~kpc$^{-3}$ along the line of sight from the Sun (left panel), along the major (positive vertical axis, middle panel) and minor axis (negative vertical axis,  middle panel) of the ellipsoidal distribution on the plane of the sky as observed from the Sun, and the distribution along the line of sight from the Milky Way centre (right panel). The panels in the middle and bottom rows show the value of $\tilde{\rho}= \rho/\rho_{\rm sph}$ and the magnitude of the internal Newtonian gravitational acceleration, respectively. In the middle and bottom rows the panels are arranged from left to right as in the top panels.  
The horizontal black lines in the bottom panels  show the magnitude of the external field, which is almost constant across the dwarf galaxy. In all the panels, the vertical black solid lines locate the centre of the dwarf. In the left and right panels, the Milky Way is at large positive radii.
}
\label{DracoDwarf}
\end{figure*}

\begin{figure*}
\centering
\includegraphics[scale=0.9]{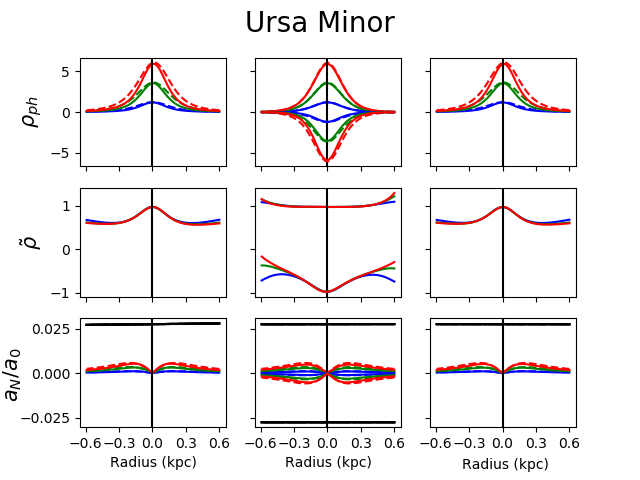}
\caption{Same as Fig.~\ref{DracoDwarf}, but for Ursa Minor.}
\label{UmiDwarf}
\end{figure*}

\begin{figure*}
\centering
\includegraphics[scale=0.9]{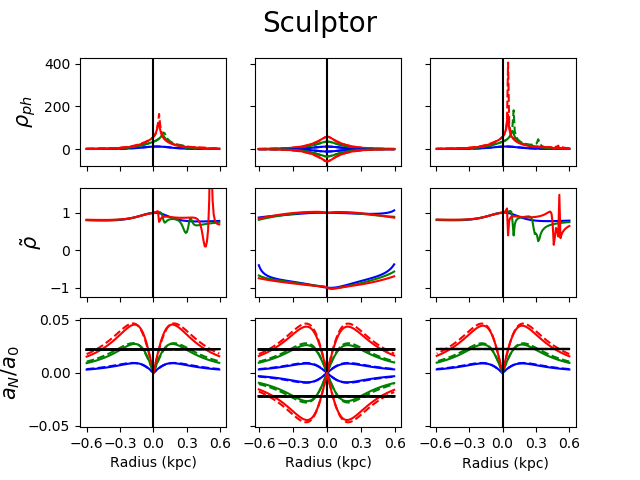}
\caption{Same as Fig.~\ref{DracoDwarf}, but for Sculptor.}
\label{SculptorDwarf}
\end{figure*}

\begin{figure*}
\centering
\includegraphics[scale=0.9]{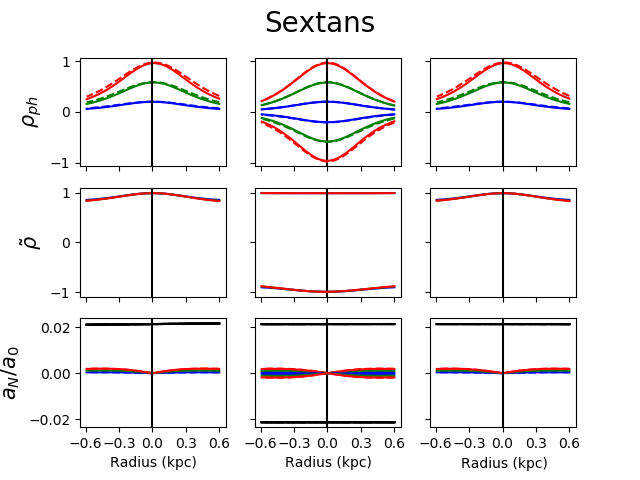}
\caption{Same as Fig.~\ref{DracoDwarf}, but for Sextans.}
\label{SextansDwarf}
\end{figure*}

\begin{figure*}
\centering
\includegraphics[scale=0.9]{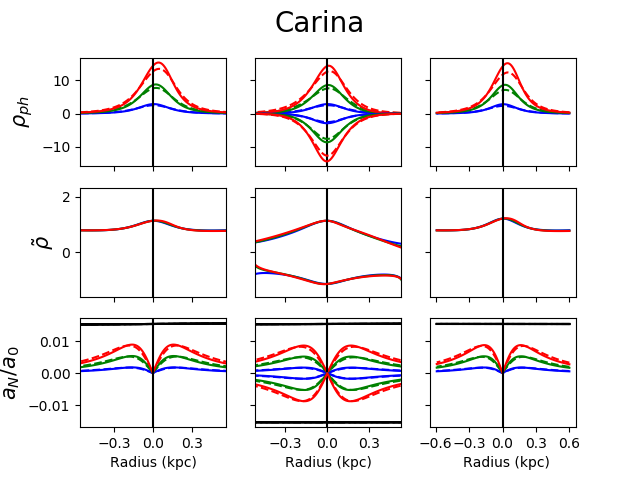}
\caption{Same as Fig.~\ref{DracoDwarf}, but for Carina.}
\label{CarinaDwarf}
\end{figure*}

\begin{figure*}
\centering
\includegraphics[scale=0.9]{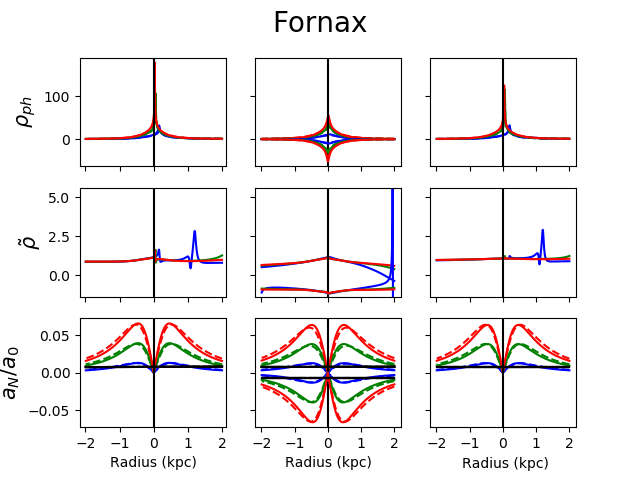}
\caption{Same as Fig.~\ref{DracoDwarf}, but for Fornax.}
\label{FornaxDwarf}
\end{figure*}

\begin{figure*}
\centering
\includegraphics[scale=0.9]{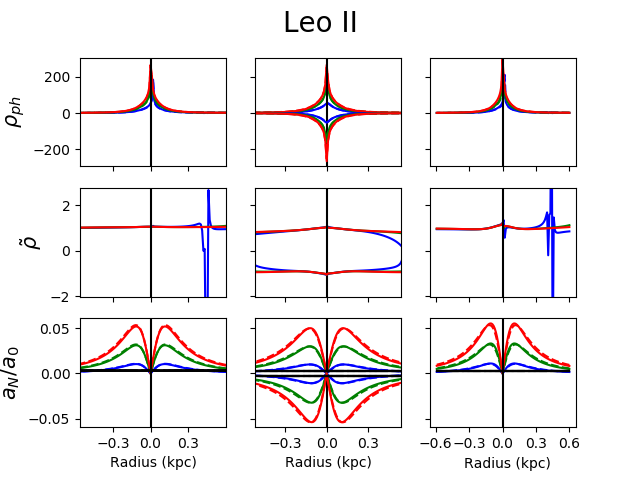}
\caption{Same as Fig.~\ref{DracoDwarf}, but for Leo II.}
\label{LeoIIDwarf}
\end{figure*}

\begin{figure*}
\centering
\includegraphics[scale=0.9]{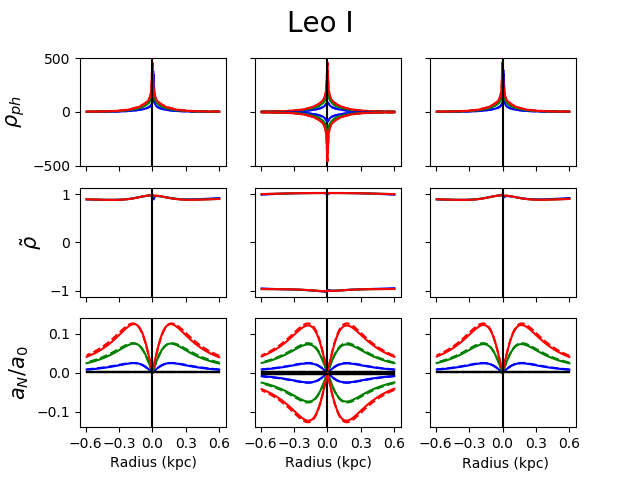}
\caption{Same as Fig.~\ref{DracoDwarf}, but for Leo I.}
\label{LeoIDwarf}
\end{figure*}

 The top panels of Figs.\ \ref{DracoDwarf} -- \ref{LeoIDwarf} show the phantom dark matter distribution for the eight dwarfs sorted by decreasing external field. The blue, green, and red curves assume a stellar mass-to-light ratio of $1$, $3,$ and $5$~$M_\odot/L_\odot$, respectively. Solid curves are for a flattened baryonic component and dashed curves are for a spherical distribution. We obtain the spherical distribution by setting $q=1$ in Eq. (\ref{rhodwarf}), and thus in turn by setting the scale radius of the sphere to the semi-major axis $b$ of the ellipsoidal distribution.

When the external field is much larger than the internal field (Carina, Draco, Sextans, and Ursa Minor), MOND predicts a Newtonian-like behaviour except for an inflated gravitational constant. When the external field is much smaller than the internal field (Leo I and Leo II) the MOND prediction returns to the isolated case. 

In the intermediate case, where the internal and external field strengths are comparable, the behaviour becomes more complicated, as anticipated in Sect.~\ref{sec:comparison} and Fig.~\ref{InternalFieldCompare}. Since the internal Newtonian gravity points towards the dwarf centre, at the far side of the dwarf from the Milky Way (i.e. at the negative axis of the radial coordinate in each of the line-of-sight panels of Figs.\ \ref{DracoDwarf}$-$\ref{LeoIDwarf}) the external field of the Milky Way acts with the internal field of the dwarf galaxy; on the other hand,  at the closest side (i.e. at the positive axis of the radial coordinate in the line of sight panels) the external field acts against the internal gravity. This difference causes the phantom dark matter predicted by MOND to be distributed non-spherically, with a density peak offset from the baryonic centre towards the Milky Way. 

The magnitude of the offset depends on the comparative strengths of the internal and external fields. In agreement with the expectation from Fig.~\ref{InternalFieldCompare}, the offset is substantial for Sculptor for any mass-to-light ratio, for Draco and Carina for high mass-to-light ratios, and for Fornax and Leo II for low mass-to-light ratios. 

 The luminosity of Ursa Minor, Carina, Draco, and Sextans are in the range $[2,4.1]\times 10^5$~L$_\odot$ and are the least luminous dwarfs in the sample (Table \ref{DwarfParams}).
Combined with their scale lengths $b$, these luminosities yield the lowest baryonic densities, and thus the weakest internal gravitational fields (Fig.~\ref{InternalFieldCompare}) and the lowest phantom densities (top panels of Figs.~\ref{DracoDwarf}-\ref{LeoIDwarf}). In Ursa Minor and Sextans, which have the lowest baryonic density, the internal gravitational field is so weak that the phantom dark matter density show a negligible offset compared to the offsets of all the remaining dwarfs.  

The direction of the EFE acceleration is crucial, as is  its magnitude. Looking at the acceleration profiles (bottom panels) for Sculptor (Fig. \ref{SculptorDwarf}), Fornax (Fig. \ref{FornaxDwarf}), and Leo II (Fig. \ref{LeoIIDwarf}), we see that the external field from the Milky Way is comparable to the internal field of the dwarf galaxy, depending on the assumed mass-to-light ratio. On the side of the dwarf closest to the Milky Way centre, these fields act in opposite directions. As we are essentially dealing with a two-body problem, there will be a point where the net Newtonian field is zero. Due to the formulation of QUMOND, at this zero point we  expect a rise in the MOND interpolation function, creating a large amount of phantom dark matter. This effect is not obvious in Figs. \ref{SculptorDwarf}, \ref{FornaxDwarf}, and \ref{LeoIIDwarf} because we are looking at the line of sight from the Sun and not from the Milky Way centre. To see this effect more clearly, the right panels of Figs. \ref{DracoDwarf} -- \ref{LeoIDwarf} show the results for when the line of sight is between the Milky Way centre and the dwarf galaxy;  
for example, for Sculptor the peak of $\rho_{\rm ph}$ for the highest mass-to-light ratio is roughly a factor of three larger compared to the profile observed from the Sun.

Along the line of sight between the Milky Way centre and the dwarf, Sculptor and Fornax (and to a lesser extent Leo II) exhibit an additional effect, namely the presence of two peaks in the phantom dark matter density profile for mass-to-light ratio $M/L=3$~$M_\odot/L_\odot$ for Sculptor and $M/L=1$~$M_\odot/L_\odot$ for Fornax and Leo II. This result is due to the shape of the internal acceleration profile. We have adopted a Plummer model \citep{Plummer1911} for the baryonic matter of the dwarf galaxy. The enclosed mass of the baryonic matter is thus proportional to  $r^3$ when $r\ll b$  and approximately constant when $r\gg b$ (with $b$ the Plummer scale length).  Therefore, as shown in the bottom panels of 
 Figs.~\ref{DracoDwarf}-\ref{LeoIDwarf}, 
the Newtonian acceleration is proportional to $r$ when $r\ll b$ and proportional to $r^{-2}$ when $r\gg b$. It follows that the internal Newtonian acceleration gets stronger as the enclosed mass grows until it reaches a maximum; it then starts to get weaker far from the object's centre. 

As shown in the bottom left and right panels of Fig.~\ref{SculptorDwarf} 
for the green line relative to the mass-to-light ratio $M/L=3$~$M_\odot/L_\odot$ for Sculptor, if the external field is slightly weaker than this maximum, there are two points where the Newtonian external field strength is equal to the Newtonian internal field strength. Therefore, if the external field is acting in the opposite direction to the internal field, there are two points where the net Newtonian gravity will tend to zero. This behaviour of the Newtonian field causes the MOND interpolation function to increase at two points resulting in two phantom dark matter density peaks. We note  that the magnitude of these effects depends on the internal Newtonian gravity, and thus on the stellar mass-to-light ratio of the dwarf.

This argument also explains the very irregular shapes in the $\tilde{\rho}$ plot. The relative strengths of the internal and external field can be different for the flattened and spherical stellar distributions, and the peak of the phantom dark matter may occur at  different places for a flattened system and a spherical system. Thus, according to the argument above, these irregular features are expected to be  more evident in Fornax and Sculptor, where the internal and external gravitational field strengths are comparable, whereas Carina, Draco, Leo I, Sextans, and Ursa Minor, where the internal field is either much weaker or much stronger than the external field, exhibit well-behaved $\tilde{\rho}$.

\subsection{Including additional external fields}
\label{sec:LeoII}

As previously mentioned, there are additional external fields acting on the dwarf galaxies in addition to the field of the Milky Way. In Sect.~\ref{LargeScaleSec}, we estimated that the additional Newtonian external field could be as strong as $a_{\rm N~ext}/a_0 \approx 0.0009$ (Eq.\ \ref{eq:aextLG}). From Fig.~\ref{InternalFieldCompare} we can see that this value is comparable to the external field of the Milky Way felt by the most distant classical dwarf galaxies, Leo II and Leo I, where the external field originating from the Milky Way is weakest. The internal field of these galaxies is larger than the external field, but there should still be an effect present. As a case study, we quantify how much the external field of the large-scale structure beyond the Local Group affects the phantom dark matter distribution of Leo II.

Following the discussion on the large-scale external fields from \cite{Wu2008}, we approximate that the external field acting on the Milky Way comes from the Great Attractor, which is in the direction of the Milky Way centre from the Sun. In our coordinate system (Sect.~\ref{sec:CoordSys})
the Great Attractor is thus located at a large negative $x$ coordinate, whereas Leo II is at the origin, and the Milky Way in  between. Therefore, the Milky Way external field and the external field from the Great Attractor will be acting together rather than opposed, and including the Great Attractor slightly increases the magnitude of the external field. 

\begin{figure}
\centering
\includegraphics[scale=0.4]{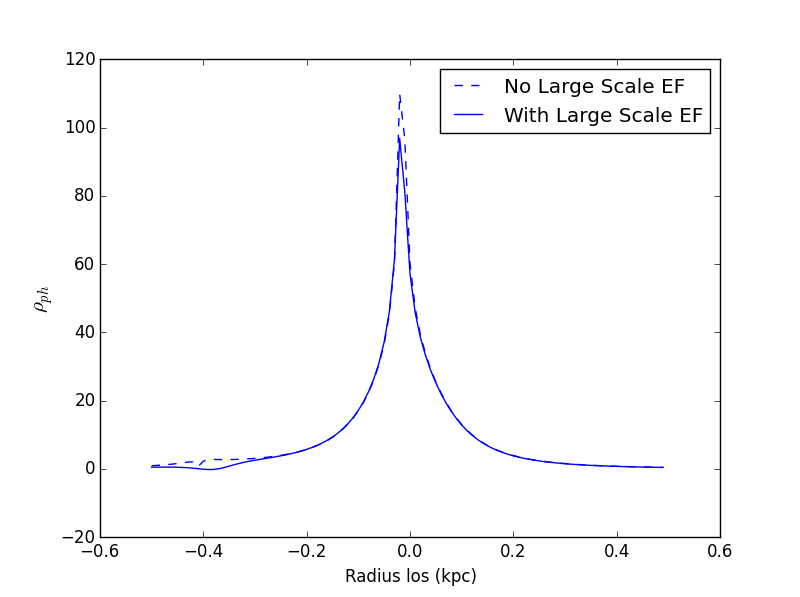}
\caption{Phantom dark matter profile in units of $10^7$~M$_\odot$~kpc$^{-3}$ for Leo II along the line of sight with and without the external field from the large-scale structure beyond the Local Group. The Milky Way is in the direction of negative $r$. See sections \ref{LargeScaleSec} and \ref{sec:LeoII} for details.}
\label{LeoII_LS_LOS}
\end{figure}

To quantify this effect, we assume a point-like mass $10^{15}$ M$_{\odot}$ at a distance of $40$ Mpc;  at this distance the external field from this source is approximately constant across the Milky Way. Figure~\ref{LeoII_LS_LOS} shows the phantom dark matter distribution along the line of sight, with $r=0$ being the dwarf centre and   negative $r$ being the side between the centre of Leo II and the centre of the Milky Way. The dotted curve is with the Milky Way external field  alone and the solid curve is with the additional external field from the Great Attractor. Here, we assume a mass-to-light ratio $M/L=1$~$M_\odot/L_\odot$ for Leo II.

Both curves are similar except at the dwarf centre  ($r=0$) and at a radius $r\sim-0.4$ kpc, the side of Leo II closer to the Milky Way. The difference at $r\sim -0.4$ kpc  arises because there is a point between the dwarf and the Milky Way where the net Newtonian gravity is zero, 
whereas the difference at the centre is due to the internal Newtonian gravity going to zero as the enclosed mass is zero at $r=0$. The interpolating function boosts the MOND acceleration at low values of   Newtonian gravity. Therefere, at both these points, $r\sim -0.4$ kpc and $r=0$, the phantom dark matter is sensitive to the small changes in the external field.

The difference we see in Fig.~\ref{LeoII_LS_LOS}  along the line of sight is the largest we can expect. The Sun is very close to the Milky Way centre, relative to the position of the dwarf, and thus the line of sight is close to the line between the Milky Way and the dwarf. It follows that  the  point where Newtonian gravity cancels out is closer to the line of sight than it is to the observed minor and major axes.  
Thus, the change in the phantom dark matter profile is greater along the line of sight than along the other two axes, which we do not show.

In conclusion, we find that the difference in the phantom dark matter distribution as a result of adding in this extra term is very small, less  than a few percentage points.

\section{Discussion}\label{DiscussionSec}

The results reported in the previous section are based on a model that goes a step forward from the usual approach adopted for dwarf
galaxies in MOND,  but it still relies on a specific choice of the interpolation function. We now briefly discuss these issues.

The middle rows of Figs.\ \ref{DracoDwarf}--\ref{LeoIDwarf} clearly show that the phantom dark matter distribution along each axis is severely altered due to the non-spherical distribution of the baryons. In principle, it might appear appropriate to model the dwarf spheroidals in MOND by solving the Jeans equations for self-gravitating axisymmetric systems, similarly to the approach of \citet{Hayashi2015} in a $\Lambda$CDM context. Unfortunately, in MOND the EFE cannot be neglected and, in general, it removes almost all the axial symmetry. However, at least for the
dwarf spheroidals where the external field dominates over the internal field, this approach might be viable because the integration of Eq.\ (\ref{PhantomDMEqn}), by ignoring the curl field, reduces to
\begin{equation}\label{MONDrescaledEqn}
\nabla(\Phi_{\rm int} + \Phi_{\rm ext}) \sim   \nu\left(\frac{|\nabla\Phi_{\rm ext~ N}|}{a_0}\right)( \nabla \Phi_{\rm N}+ \nabla\Phi_{\rm ext~N}) \, , 
\end{equation}
 and the MOND gravitational field becomes $ a = a_{\rm N}\times G_{\rm eff}/G$, where  $a_{\rm N} = -\nabla \Phi_{\rm N}- \nabla\Phi_{\rm ext~N}$ is the Newtonian gravity and $G_{\rm eff}=G\nu(\vert\nabla\Phi_{\rm ext~ N}\vert/a_0)$ is a re-scaled gravitational constant.
Therefore, although these galaxies can be properly described by a mass-follows-light model \cite[e.g.][]{Lokas2009}, when interpreted in Newtonian gravity they are still dominated by massive halos of dark matter \citep[e.g.][]{Walker2009}.

For the remaining dwarfs, the approach we present in this work is the simplest that returns reliable results.
Our approach is also   to be preferred to the  approximation sometime adopted to study systems in QUMOND, when a constant external field is present.   This approximation derives from the following argument. The integration of Eq.\ (\ref{PhantomDMEqn}),   ignoring again the curl field, gives the MOND internal field,   
\begin{align}\label{MONDapprox1}
\begin{split}
        \nabla \Phi_{\rm int}(r) &\simeq \nabla\Phi_{\rm N}(r) \nu \left( \frac{|\nabla\Phi_{\rm N}(r) + \nabla\Phi_{\rm N~ext}|}{a_0} \right) \\ &+ \nabla\Phi_{\rm N~ext} \left[  \nu \left( \frac{|\nabla\Phi_{\rm N}(r) + \nabla\Phi_{\rm N~ext}|}{a_0}\right) -  \nu \left( \frac{| \nabla\Phi_{\rm N~ext}|}{a_0}\right) \right] \, ,
\end{split}
\end{align}
where the MOND external field is estimated as $\nabla\Phi_{\rm ext}\simeq \nu ( \vert \nabla\Phi_{\rm N~ext}\vert /a_0) \nabla\Phi_{\rm N~ext}$. For external fields  generated by  
flat systems like the Milky Way, $\nabla\Phi_{\rm N~ext}$ has a preferred direction 
and it can be ignored when considering directions perpendicular to it. In this case, the second term on the right-hand side of Eq.\ (\ref{MONDapprox1}) can be neglected, and we have
\begin{equation}\label{MONDapprox3}
        \nabla \Phi_{\rm int}(r) \approx \nabla\Phi_{\rm N}(r) \nu \left( \frac{\sqrt{\vert \nabla\Phi_{\rm N}(r)\vert^2 + \vert \nabla\Phi_{\rm N~ext }\vert^2}}{a_0} \right) 
,\end{equation}
where the approximation
\begin{equation}\label{MONDapprox2}
|\nabla\Phi_{\rm N}(r) + \nabla\Phi_{\rm N~ext}| \approx \sqrt{\vert\nabla\Phi_{\rm N}(r)\vert^2 + \vert\nabla\Phi_{\rm N~ext }\vert^2}
\end{equation}
again holds in the direction  
perpendicular to the external field \citep[e.g.][]{Haghi2016}. 
Although this approximation is not completely accurate, as it neglects the specific direction of the external field with respect to the internal field \citep[e.g.][]{Famaey2012}, Eq.\ (\ref{MONDapprox3}) is often acceptable for first-order approximations \citep[e.g.][]{Angus2008}; moreover, when $\vert \nabla\Phi_{\rm N}\vert \gg \vert \nabla\Phi_{\rm N~ext }\vert$, Eq.\ (\ref{MONDapprox3}) reduces to 
\begin{equation}\label{MONDapprox4}
        \nabla \Phi_{\rm int}(r) \approx \nabla\Phi_{\rm N}(r) \nu \left( \frac{\vert\nabla\Phi_{\rm N}(r)\vert }{a_0} \right)\; ,
\end{equation}
which is the standard formula for spherical systems. 

Our results in the previous section, however, highlight the need to move beyond this first-order approximation as phantom dark matter offsets will be missed if spherical symmetry is assumed. %\LEt{ see note 2  }
Our full approach  shows that for some dwarf spheroidals MOND can predict a phantom dark matter peak significantly offset from the baryons. This effect   is most prominent in the case of Fornax and Sculptor, where the internal and external fields are comparable and the approximations  of   Eq.\ (\ref{MONDrescaledEqn}) and Eq.\ (\ref{MONDapprox4}) do not apply. The position of the peaks depends on the relative strength of the internal and external fields. In the case of Fornax and Sculptor the peak appears close to 100 pc or even 200 pc from the centre in the direction of the external field. This displacement is not trivial because it is of the order of the dwarf half-mass radius, which is $\sim 1.3$ times the scale radius $b$ (see Table \ref{DwarfParams}). The exact value of the offset depends on the stellar mass-to-light ratio used. 

Therefore, if the data for a dwarf spheroidal interpreted in a Newtonian framework suggested  
a dark component that is aligned with the baryons, we would conclude that $\Lambda$CDM should be preferred over modified gravity. However, the opposite is not trivially true. 
After analysing the EAGLE \citep{Schaye2015} and cosmo-OWLS \citep[][]{LeBrun2014} simulations, \citet[][]{Velliscig2015} show that the stellar components can in principle be misaligned from the host halo in $\Lambda$CDM. However, detailed numerical simulations of dwarf galaxies would need to be conducted to quantify the largest displacements between dark halos and their stellar component allowed in the standard model.

We are also  assuming a static set-up. In reality the dwarf is in motion around the Milky Way with the external field constantly changing in time \citep{Angus2014}. The density and velocity fields of the stars within the dwarf do not instantaneously adapt to the varying external field and the
observables might thus mirror a situation slightly different from the situation expected at the current position of the dwarf in a static model \citep{Famaey2018}. Numerical simulations could potentially quantify this dynamical delay and quantify, in a more realistic scenario, the phantom dark matter displacements,   although  we do not expect them to be substantially different from those suggested here.

 From these more realistic models, we could derive specific predictions on the distortion of the density and velocity fields of the stars within the dwarf. These distortions could, in principle, be observable with accurate measurements of the surface brightness distribution of the dwarf and with measurements of the proper motions of the dwarf stars with future astrometric missions \citep{Theia2017, Theia2019}. Specifying these theoretical predictions and the observational sensitivity to validate them requires further investigations.

We conclude our discussion with a note on the singularity of the interpolation function when the Newtonian gravitational field is zero. Singularities often appear in toy models like ours. However, the qualitative prediction remains valid. For example, the \citet{Navarro1996} density profile is singular at the origin, but a cuspy density profile of dark matter halos remains a general prediction of the standard model. Similarly, although the singularity of the interpolation function is indeed responsible for the spikes of the phantom dark matter distributions in Fornax and Sculptor, we expect that moving to a more realistic interpolation function will reduce, but not remove, these spikes. Due to these singularities, when plotting the phantom dark matter density profiles the particular choice of discretisation and step size sets the position of the points around the singularity; therefore, the lines around these points can vary slightly. However, away from these singular points, the general features of the phantom dark matter density profiles remain unchanged, even when slightly larger or smaller step sizes are used.

There are MOND-like formulations that do not require an interpolation function, such as the one presented in \cite{Babichev2011} whose cosmological model has been investigated by \cite{Zlosnik2017}. In this model, which is based on the Tensor-Vector-Scalar (TeVeS) formulation of Bekenstein \citep{Bekenstein2004}, the total gravitational field is composed of a Newtonian term and a scalar field. In TeVeS the scalar field is controlled by an interpolation function; instead, the model of \cite{Babichev2011} suppresses the scalar field using a Vainshtein screening term, usually found in Galileon 
theories of dark energy. Unfortunately, on non-linear scales this model has not yet been explored beyond spherical symmetry, and therefore we cannot apply it to our systems to explore its predictions when the Newtonian field goes to zero.

\section{Conclusion}\label{ConclusionsSec}

We have computed the phantom dark matter density predicted in MOND for eight dwarf spheroidal galaxies in the vicinity of the Milky Way. By approximating the stellar distribution within the dwarf spheroidals with an axisymmetric flattened spheroid, we show how  (1) the aspherical shape of the baryons and (2) the external gravitational field affect the phantom dark matter distribution in these objects. 

As expected, the flattening of the stellar component yields significantly altered phantom dark matter distributions compared to the spherical case. The level of flattening of the phantom dark matter density depends on both the flattening of the stellar component and the direction of the external field.

The EFE is a major contributor in determining the distribution of the phantom dark matter in the dwarf spheroidals. In most dwarfs, Draco, Sculptor, Carina, Fornax, and Leo II, the phantom dark matter is offset from the baryons by a substantial fraction of the half-mass radius. In addition, for Sculptor and Fornax, where the internal and external gravitational fields are comparable, the phantom dark matter distribution displays substantial spikes where the two gravitational fields cancel each other; these spikes are basically unrelated to the stellar distribution.  
For the second most distant dwarf spheroidal in our sample, Leo II, the large-scale external field  beyond the Local Group has an additional small effect on the phantom dark matter distribution, which clearly depends on the exact external field strength and direction. 

The offset in the phantom dark matter density profile in most dwarfs and the spikes in Sculptor and Fornax are a genuine prediction of MOND and could potentially discriminate between MOND and the standard $\Lambda$CDM model. 
For example, dwarf spheroidals might be able to help constrain the gravitational potential of the Milky Way and also the large-scale external field exerted on the Milky Way: each dwarf lies at a different position from the Milky Way; therefore, by comparing the dynamics of multiple dwarf spheroidals one may be able to argue for or against MOND. There are pairs of dwarf galaxies that lie at similar distances (within a few parsecs) from the Milky Way centre: Leo I and Leo II, Ursa Minor and Draco, Sextans and Sculptor. If these dwarf spheroidals require  significantly different external fields to explain their dynamics, this could present a challenge to MOND. On the other hand, if the MOND external field effect describes the dynamics in a self-consistent manner, this would be a very strong case for MOND.

As a cautionary tale, we note that if the phantom dark matter offsets we find here are not detected, we cannot conclude that MOND is ruled out. The offsets are governed by the interaction between the external and the internal fields. There are MOND-like ideas, such as Modified Inertia  \citep{Milgrom1994}, where the EFE behaves in a different manner to that of the more familiar 
MOND formulation we adopt here \citep{Milgrom2011}. Therefore, in principle, it is not possible to conclude that MOND as a concept is ruled out if such offsets are not observed, but these observations could provide insight into the exact formulation of MOND. 

We finally acknowledge that we are working with a very simplified model where the baryons are distributed according to analytical models. We are also assuming a static potential. These assumptions are indeed usual and are often applied to dwarf spheroidals in a $\Lambda$CDM context \citep[e.g.][]{Walker2009},  
although  numerical simulations in MOND can  more realistically model a time-dependent EFE affecting the dwarfs on their orbits around the Milky Way \citep{Angus2014}. Our goal here was to highlight some interesting features that arise when applying these same assumptions to a MOND framework. Due to the non-linear nature of MOND, there are some nuances that should be carefully considered compared to a $\Lambda$CDM framework. 

We believe that highlighting these issues will eventually lead to more detailed predictions of dwarf spheroidal dynamics, and ideally lead to an observational distinction between $\Lambda$CDM and MOND.
For example, the inclusion of proper motions of the stars in dwarf spheroidals, in addition to their velocity dispersion profiles, when modelling their dynamics was proposed in the past \citep[e.g.][]{Wilkinson2002,Strigari2007} and  by a future astrometric Theia-like mission \citep{Theia2017, Theia2019}, and was recently achieved for Sculptor, based on Gaia data \citep{Massari2018}. The proper motions of stars can vastly improve the constraints on the internal dynamics of dwarfs \citep[e.g.][]{Strigari2018a} and thus provide additional stringent tests of the theory or gravity. We plan to investigate the role of proper motions in this context in future work.

%%% Acknowledgements
\begin{acknowledgements}
We sincerely thank the referee for constructive suggestions and Matthew Walker, Kohei Hayashi, and Mike Irwin for useful discussions on dwarf spheroidals. AH was supported by Progetto di Ateneo/Compagnia di S. Paolo 2016, {\it The Milky Way and Dwarf Weights with Space Scales}. We also acknowledge partial support from the Italian Ministry of Education, University and Research (MIUR) under the {\it Departments of Excellence} grant L.232/2016, and from the INFN grant InDark. 
 This research has made use of NASA's Astrophysics Data System Bibliographic Services.
\end{acknowledgements}

%%%%%%%%%%%%%%%%%%%%%%%%%%%%%%%%%%%%%%%%%%%%%%%%%%

%%%%%%%%%%%%%%%%%%%% REFERENCES %%%%%%%%%%%%%%%%%%

\bibliographystyle{aa} % style aa.bst
\bibliography{QUMONDbib} % your references Yourfile.bib

\end{document}